# Formation and Signatures of the First Stars [†]

Z. Haiman & A. Loeb

*Harvard University, 60 Garden Street, Cambridge,*
*MA 02138, USA*

We use a spherical hydrodynamics code to show that in cold dark matter cosmologies, the first stars form at $z \sim 50$ through the direct collapse of gas in low–mass systems ($\sim 10^4 M_\odot$). Photons from the first stars easily photodissociate $H_2$ throughout the universe and so molecular cooling does not affect the subsequent fragmentation of gas clouds into stars. We examine observable signatures of the pre–galactic population of stars. These include the detected metallicity and photo-ionization of the intergalactic medium, and the soon to be detected damping of microwave anisotropies on small angular scales ($\lesssim 10°$). The Next Generation Space Telescope will be able to directly image the pre–galactic star clusters, while the DIMES experiment could detect their Bremsstrahlung emission.

In popular Cold Dark Matter (CDM) cosmologies, the first baryonic objects form at redshifts as high as $z \sim 50$. Although these redshifts are well beyond the current horizon of direct observations, $z \sim 5$, the existence of pre–galactic stars has a number of observable consequences. The stars could reionize the intergalactic medium (in accordance with the lack of the Gunn–Peterson effect out to $z \sim 5$), and the resulting optical depth to electron scattering would damp the microwave background anisotropies on angular scales $\lesssim 10°$. The latter signature will be searched for by future satellite experiments such as MAP or COBRAS/SAMBA. Pre–galactic stars could also produce the amount of carbon necessary to explain the roughly universal $\sim 1\%$ solar metallicity that was detected recently [1,2] in Ly$\alpha$ absorption systems at $z \sim 3$. Finally, an old population of low–mass ($M \lesssim 3 M_\odot$) stars would behave as collisionless matter during galaxy formation and populate the diffuse halos of galaxies. Such stars might account for some of the events observed by ongoing microlensing searches in the halo of the Milky Way [3] and could also be detected in the future through their lensing of distant quasars [4].

We have quantified these observational signatures in CDM cosmologies [5] using a simple semi–analytic approach in which clouds virialize according to the Press–Schechter theory and fragment into stars with an efficiency that yields the observed C/H ratio. This approach is complimentary to more detailed, but computationally expensive and less versatile 3–D numerical simulations [6]. The main ingredients of our model are:

(i) *The Collapsed Fraction of Baryons.* We use the Press–Schechter theory to find the abundance and mass distribution of virialized dark matter halos.

---

[†] To appear in the Proceedings of the 18th Texas Symposium

However, the collapse of the baryons is delayed relative to the dark matter in low mass objects where gas pressure force resists gravity. We obtained the exact collapse redshifts of spherically symmetric perturbations by following the motion of both the baryonic and the dark matter shells with a one dimensional hydrodynamics code [7]. We find that due to shell–crossing with the cold dark matter, baryonic objects with masses $10^{2-3} M_\odot$, well below the linear–regime Jeans mass, are able to collapse by $z \sim 10$. In calculating the abundance of virialized objects, we take into account the increase in the effective Jeans mass due to photoionization.

(ii) *Star Formation.* We calibrate the fraction of the gas which is converted into stars in each virialized cloud based on the inferred C/H ratio in the Ly$\alpha$ absorption forest. We use tabulated $^{12}$C yields of stars with various masses [8], and consider three different initial mass functions (IMFs). In addition, we include a negative feedback on star–formation due to the photodissociation of molecular hydrogen by photons with energies in the range 11–13.6 eV. We find that soon after the appearance of the first few stars, molecular cooling is suppressed even inside dense objects [9]. Due to the lack of any other cooling agent in the metal–poor primordial gas, the bulk of the pre–galactic stars form due to atomic line cooling and fragmentation inside massive objects ($\gtrsim 10^8 M_\odot$) with virial temperatures $\gtrsim 10^4$K.

(iii) *Propagation of Ionization Fronts.* The composite spectrum of ionizing radiation which emerges from the star clusters is determined by the stellar IMF and the recombination rate inside the cluster. We follow the time-dependent spectrum of a star of a given mass based on standard spectral atlases [10] and the evolution of the star on the H–R diagram as prescribed by theoretical evolutionary tracks [11]. To calculate the fraction of the ionizing photons lost to recombinations inside their parent clouds, we adopt the equilibrium density profile of gas inside each cloud according to our spherically–symmetric simulations. We then use the time–dependent composite luminosity of each star-forming region to calculate the propagation of a spherical ionization fronts into the surrounding IGM.

Table 1 summarizes our results for a range of parameters. We varied the cosmological power spectrum ($\sigma_{8h^{-1}}$, $n$), the baryon density ($\Omega_{\rm b}$), the star formation efficiency ($f_{\rm star}$), the escape fraction of ionizing photons ($f_{\rm esc}$), the IMF, and whether or not the negative feedback due to $H_2$ is present. For almost the entire range of parameters, the universe is reionized by a redshift $\gtrsim 10$. The optical depth to electron scattering is in the range $\sim 0.1$–$0.2$. The resulting damping of $\sim 10$–$20\%$ for the amplitude of microwave anisotropies will be detectable with the MAP or the COBRAS/SAMBA satellites. The only exception occurs when the IMF is strongly tilted towards low–mass stars (by a power law index of 1.7), and reionization is suppressed due to the absence of massive stars which ordinarily dominate the ionizing flux. In this case, the

Table 1: Reionization redshift and electron scattering optical depth for a range of parameters.

| Parameter | Standard | Range Considered | $z_{\rm reion}$ | $\tau_{\rm e.s.}$ |
|---|---|---|---|---|
| $\sigma_8 h^{-1}$ | 0.67 | 0.67–1.0 | 25–32 | 0.14–0.20 |
| $n$ | 1.0 | 0.8–1.0 | 18–25 | 0.08–0.14 |
| $\Omega_{\rm b}$ | 0.05 | 0.01–0.1 | 24–30 | 0.03–0.26 |
| $f_{\rm star}$ | 4% | 4%–90% | 20–35 | 0.10–0.20 |
| $f_{\rm esc}$ | $f_{\rm esc}(z)$ | 3%–100% | 14–31 | 0.08–0.18 |
| IMF tilt ($\beta$) | 0 | 0–1.69 | <25 | 0.01–0.14 |
| $H_2$ feedback | yes | yes/no | 25–29 | 0.14–0.19 |

large density of low–mass stars could account for the MACHO events detected towards the LMC. [3]

Another detectable signature of reionization is free–free emission. We find[5] that the contribution of free–free emission from star–forming clouds at $z > 10$ to the brightness temperature of the microwave sky is well above the proposed sensitivity of the DIMES experiment[12] in the frequency range $\nu = 1$–100 GHz.

Finally, the Next Generation Space Telescope[13] will be able to directly image and resolve the high redshift star clusters with its projected sensitivity of $\sim 1$ nJy in the range 1–3.5$\mu$m, and angular resolution of 0.06″. Based on our calculated luminosity function of star clusters, we find[5] that NGST will probe $\sim 10^4$ objects per field of view at $z \gtrsim 5$.

We thank P. Höflich, M. Rees, and D. Sasselov for useful discussions.